\documentclass[twocolumn,aps,superscriptaddress,nofootinbib,floatfix]{revtex4}
\usepackage{lineno}
\usepackage{float}
\usepackage{listings}
\usepackage[utf8]{inputenc}
\usepackage{hyperref}
\usepackage{amsmath}
\usepackage{amssymb}
\usepackage{amsfonts}
\usepackage{tabularx}
\usepackage{booktabs}
\usepackage{graphicx}
\usepackage{subfigure}
\usepackage{color}
\usepackage{diagbox}
\usepackage{multirow}
\usepackage{bm}
\usepackage[inline]{enumitem}
\usepackage[capitalize]{cleveref}
\usepackage[T2A,T1]{fontenc}
\usepackage[title]{appendix}
\graphicspath{{fig}}
\definecolor{theblue}{RGB}{0,50,230}

\usepackage{hyperref}
\hypersetup{
	colorlinks=true,
	linkcolor=theblue,
	citecolor=theblue,
	urlcolor=theblue
}

\newcommand {\avg}[1]{\ensuremath{\langle\kern-1.0pt\langle#1\rangle\kern-1.0pt\rangle}}

\def\eq{{\,=\,}}

\newlength\cmsFigWidth

\def\pb{\phi,b}
\def\ppb{p_T,\phi,b}
\def\pt{p_T}
\def\bq{\begin{eqnarray}}
	\def\eq{\end{eqnarray}}

\usepackage[normalem]{ulem}  

\renewcommand\sout{\bgroup \color{red} \ULdepth=-.5ex \ULset}

\begin{document}
	
	\title{Spectra and azimuthal harmonic of charmed mesons in Pb+Pb collisions at LHC energies}
	
\author{Jingzong Zhang}
\affiliation{College of Physics, Sichuan University, Chengdu 610064, China}
\author{Huanjing Gong}	
\affiliation{College of Physics, Sichuan University, Chengdu 610064, China}
\author{Hua Zheng}\email{zhengh@snnu.edu.cn}
\affiliation{School of Physics and Information Technology, Shaanxi Normal University, Xi'an 710119, China}
\author{Wei Dai}
\affiliation{School of Mathematics and Physics, China University of Geosciences, Wuhan 430074, China}
\author{Lilin Zhu}\email{zhulilin@scu.edu.cn}
\affiliation{College of Physics, Sichuan University, Chengdu 610064, China}

\date{\today}
	
\begin{abstract}
Using the recombination model (RM) extended to the heavy flavor scenario, we investigate the production of charmed mesons, i.e., $D^0$, $D_s^+$ and $J/\psi$, in Pb+Pb collisions at $\sqrt{s_{NN}}=2.76$ and 5.02 TeV at the CERN Large Hadron Collider (LHC) including the heavy flavor energy loss inside the hot and dense medium. Soft, semihard, and hard partons all play important roles in our model and are uniformly treated for all mesons produced. The resulting transverse momentum spectra and the second azimuthal harmonic $v_2$ for these mesons are presented and compared with the available experimental data. The good agreement provides strong support again that the quark recombination mechanism is a universal scheme of hadronization.

\end{abstract}
	
\maketitle

\section{introduction}
The investigation of strongly interacting matter under extreme conditions of temperature and energy density is at the forefront of nuclear physics research, which is a formidable challenge to map out the phase structure of quantum chromodynamics (QCD) \cite{Busza:2018rrf, Borsanyi:2010bp, Bazavov:2011nk}. The experiments of high energy heavy-ion collisions at the BNL Relativistic Heavy Ion Collider (RHIC) and CERN Large Hadron Collider (LHC) provide a unique opportunity to create the hot dense QCD matter and study its properties. Up to now, a fundamental question that remains open is how the quark-hadron emerges from the underlying in-medium QCD force. The heavy quarks play a special role in this field \cite{Averbeck:2013oga, Rapp:2009my}. Since the masses of heavy quarks are much higher than the pseudo-critical temperature, they retain through the hadronization transition in ultra-relativistic heavy-ion collisions. In addition, the energy loss is predicted to depend on the quark mass and to be smaller for heavy quarks \cite{vanHees:2005wb}. Therefore, the hadrons containing charm quarks ($c$) or their antiquarks ($\bar c$) are excellent probes of the hadronization mechanism. 

The production of charmed mesons in heavy-ion collisions have been widely experimentally measured at RHIC \cite{STAR:2018zdy, STAR:2017kkh, STAR:2013eve} and LHC \cite{ALICE:2014qvj, ALICE:2013olq, ALICE:2013xna, ALICE:2015vxz, ALICE:2015dry, ALICE:2015jrl, ALICE:2021rxa, ALICE:2021kfc, ALICE:2023gco, ALICE:2017pbx, ALICE:2017quq}. To understand these results, several theoretical groups have developed transport approaches to describe the dynamics of heavy quarks in their productions at the initial state, their interactions with the quark gluon plasma (QGP), the hadronization process, and the interactions of heavy hadrons in the hadronic phase.
In those models, the hadronization of heavy quarks is modeled with fragmentation or a hybrid fragmentation plus recombination scenario ~\cite{vanHees:2005wb,He:2011qa,Minissale:2020bif,Cao:2015hia,Cao:2016gvr, Cao:2018ews,Cao:2019iqs,Gossiaux:2009mk,Song:2015sfa,Song:2015ykw,He:2019vgs,Li:2020zbk,Beraudo:2022dpz, Braaten:1994bz, Andronic:2015wma, Peterson:1982ak, Cacciari:2005rk, Sjostrand:2006za}. Heavy flavor quarks are directly produced in the hard processes of p+p collisions, providing a cleaner probe for investigating the in-medium hadronization mechanism, not only for open mesons but also shedding light on the production of hidden charm mesons in high-energy nuclear-nuclear collisions. In our previous works, we studied the light hadron production in Au+Au collisions at $\sqrt{s_{NN}}=7.7-200$ GeV and Pb+Pb collisions at $\sqrt{s_{NN}}=2.76$ and 5.02 TeV \cite{hz2, Zhu:2014csa, Zhu:2021fbs}. It established that the recombination model can nicely describe the light hadron production, which includes the effects of minijets. In particular, we have found compact formulas to describe the parton momentum distributions at the surface of the medium at midrapidity for any centrality, after the hard and semihard partons undergo momentum degradation when transversing the medium. 
Therefore, it is worth extending our study to the open- and hidden-charm mesons. In this paper, we will employ the quark recombination model to investigate both the transverse momentum spectra and the second azimuthal harmonic coefficient of $D^0$, $D_s^+$ and $J/\psi$ produced in Pb+Pb collisions at $\sqrt{s_{NN}}=2.76$ and 5.02 TeV, respectively.  

This paper is organized as follows: Section \ref{RM} briefly introduces the basic framework of the recombination model and the setups of the calculations. The formalisms of parton recombination for $D^0$, $D_s^+$ and $J/\psi$ are shown in Section \ref{DIS}. In Sec. \ref{results}, we show the results from our study on the centrality dependence of transverse momentum spectra and the second harmonic coefficient $v_2$ of $D^0$, $D_s^+$ and $J/\psi$ at midrapidity in Pb+Pb collisions at $\sqrt{s_{NN}}=2.76$ and 5.02 TeV. Finally, Sec. \ref{summary} summarizes the results and gives the conclusion from the present study.

\section{recombination of charm quark}\label{RM}
To understand the transverse momentum distributions of charmed mesons, one can consider the 1D formulation of recombination given in Refs. \cite{hy4, Hwa:2018qss, hz2, Zhu:2014csa, Zhu:2021fbs}. The invariant distribution of mesons is expressed as
\begin{eqnarray}
p^0{dN^M \over dp_T}=\int {dp_1 \over p_1}{dp_2 \over p_2}F_{q_1 \bar{q_2}}(p_1,p_2)R^M_{q_1 \bar{q_2}}(p_1,p_2,p_T), \label{21}
\end{eqnarray}
where $R^M_{q_1 \bar{q_2}}$ is the recombination function (RF) for the formation of a meson with the transverse momentum $p_T$ composed of a quark $q_1$ with momentum $p_1$ and an antiquark $\bar{q_2}$ with momentum $p_2$, 
\begin{eqnarray}
R^M(p_1,p_2,p)&=&{g_M \over B(a,b)}\nonumber\\
	&&\times ({p_1 \over p})^{a} ({p_2 \over p})^{b} \delta({p_1 \over p}+{p_1 \over p}-1),
\label{22}
\end{eqnarray}
where $g_M$ is the statistical factor and $B(a, b)$ is the beta function. 
Having determined the RF, the natural question next is how to determine the parton distribution $F_{q_1 \bar{q_2}}$ just before hadronization. In the recombination model, the partons are divided into two types: thermal (T) and shower (S). The thermal partons contain the hot medium effect, while the shower partons are due to semihard and hard scattered partons. For mesons, the partons have the following recombination types,
\begin{eqnarray}
F_{q_1 \bar{q_2}}(p_1,p_2)=\mathcal{TT}+\mathcal{TS}+\mathcal{SS},
\label{23}
\end{eqnarray}
where $\mathcal{T}$ and $\mathcal{S}$ are the thermal and shower partons invariant distributions, respectively. 

For the thermal parton distribution, a simple exponential form is assumed, 
\begin{eqnarray}
\mathcal{T}_j=p_1{dN_j^T \over dp_1}=C_jp_1 e^{-p_1/T_j},
\label{24}
\end{eqnarray}
where $C_j$ has the dimension of inverse momentum and $T_j$ refers to the inverse slope. We use subscripts $j=\{q, s, c\}$ corresponding to light ($u, d$), strange, charm quark, respectively. The values of $T_q$ and $T_s$ for Pb+Pb collisions at $\sqrt{s_{NN}}=2.76$ and 5.02 TeV have been obtained in our earlier work \cite{Zhu:2021fbs} and listed in TABLE \ref{tab1}, while $T_c$ will be discussed in Sec. \ref{results}.
 
 \begin{table}
\tabcolsep0.3in
\renewcommand{\arraystretch}{1.3}
\begin{tabular}{ccc}
\hline
\hline
 $\sqrt{s_{NN}}$ (TeV) &2.76 &5.02\\ 
 \hline
$T_q$ (GeV) & 0.39 & 0.415 \\
$T_s$ (GeV) &0.51 &0.545 \\
$T_c$ (GeV) &0.57 &0.79\\
 \hline
  \hline
 \end{tabular}
 \caption{Parameters $T_q$, $T_s$ and $T_c$ for Pb+Pb collisions at $\sqrt{s_{NN}}=2.76$ and 5.02 TeV, respectively.} 
 \label{tab1}
 \end{table}
 
The consideration of shower partons is a unique feature of the recombination model, which is empowered by the possibility to include fragmentation process as $\mathcal{SS}$ recombination for mesons. At a given centrality $c$, the shower parton distribution after integration over jet momentum $q$ and summed over all jets is defined as \cite{Hwa:2018qss, Zhu:2014csa, Zhu:2021fbs}
 \begin{eqnarray}
 \mathcal{S}^j(p, c)=\int{dq\over q} \sum_{i} \hat{F}_i(q, c) S_i^j(p, q).
 \label{25}
\end{eqnarray}
$S_i^j(p, q)$ is the unintegrated shower-parton distribution (SPD) in a jet of type $i$ fragmentating into a parton of type $j$ with momentum fraction $p/q$. It is determined by the fragmentation function (FF) on the basis that hadrons in a jet are formed by recombination of the shower partons in the same jet \cite{hy4, Peng:2010zza}. $\hat{F}_i(q, c)$ is the distribution of hard or semihard parton $i$ at the medium surface after momentum degradation while transversing the medium but before fragmentation,
\begin{eqnarray}
\hat F_i(q, c)&=&\frac{1}{2\pi}\int d\phi\int d\xi P_i(\xi, \phi, c) \\ \nonumber
&&\times\int dkkf_i(k, c)G(k, q, \xi),
\label{26}
\end{eqnarray}
where $G(k, q, \xi)$ is the momentum degradation function  due to energy loss. As discussed in Ref. \cite{Hwa:2018qss}, it was parametrized as
\begin{eqnarray}
G(k, q, \xi)=q\delta(q-ke^{-\xi}). \label{27}
\end{eqnarray}
$P_i(\xi, \phi, c)$ is the probability for parton $i$ having a dynamical path length $\xi$ at angle $\phi$ initiated at position $(x_0, y_0)$, weighted by the nuclear overlap function and integrated over all $(x_0, y_0)$. The dynamical path length $\xi$ is proportional to the geometrical path length $l$, which is calculable from nuclear geometry \cite{Chiu:2008ht}. Therefore, $P_i(\xi, \phi, c)$ can be written as
\begin{eqnarray}
P_i(\xi, \phi, c) =\int dx_0dy_0Q(x_0, y_0, c)\delta(\xi-\gamma_i l(x_0, y_0, \phi, c)),
\label{28}
\end{eqnarray}
where $Q(x_0, y_0, c)$ is the probability that a hard (or semihard) parton is produced at $(x_0, y_0)$, which can be calculated from nuclear thickness functions \cite{hz2, Hwa:2009tx}. The factor $\gamma_i$ is introduced to account for the effects of jet quenching in the medium that results in additional parton degradation due to the soft partons created. Due to many more minijets produced and the different geometrical configuration of colliding system, the $\gamma_i$ factor at LHC is different from that at RHIC . For Pb+Pb collisions, we parametrized $\gamma_g$ for gluons as \cite{Zhu:2014csa}
\begin{eqnarray}
\gamma_g(q)=\frac{\gamma_0}{1+(q/q_0)^2}.
\label{29}
\end{eqnarray}
The parameters $\gamma_0$ and $q_0$ are determined by fitting the spectra in the intermediate $p_T$ region. As discussed before, the gluons lose about twice as much energy as light quarks ($u, d, s$) \cite{Zhu:2014csa}, which directly implies that gluons on average lose the same fraction of momentum as quarks do in half distance of transversal through the hot dense medium. The earlier work has shown that it is an important factor to reproduce the light hadron spectra \cite{Zhu:2014csa}. The values of the two parameters for gluons for Pb+Pb collisions at 2.76 TeV and 5.02 TeV are listed in TABLE \ref{tab2} \cite{Zhu:2014csa, Zhu:2021fbs}. Their values for charm quarks will be discussed in Sec. \ref{results}.

\begin{table}
\tabcolsep0.25in
\renewcommand{\arraystretch}{1.3}
\begin{tabular}{ccc}
\hline
\hline
 $\sqrt{s_{NN}}$ (TeV) &2.76 &5.02\\ 
 \hline
$\gamma_0$  & 2.8 & 4.5 \\
$q_0$ (GeV/c) &7 &7 \\
 \hline
  \hline
 \end{tabular}
 \caption{Parameters $\gamma_0$ and $q_0$ for $\gamma_g$ in Eq. (9) for Pb+Pb collisions at $\sqrt{s_{NN}}=2.76$ and 5.02 TeV, respectively.} 
 \label{tab2}
 \end{table}
 
The minijets generate shower partons after emerging from the medium surface. Those shower partons recombine with themselves or with thermal partons in various combinations to form hadrons. The transverse momentum distribution of minijets $f_i(k, c)$ contains the shadowing effect of the parton distribution in nuclear collisions. For gluons, light quarks $(i=u, d, s)$ and their antiquarks at central collisions, a simple parametrization is given as follows \cite{sgf}
\begin{eqnarray}
f_i(k)=K\frac{A}{(1+k/B)^{\beta}},
\label{210}
\end{eqnarray}
where $K=2.5$. For Pb+Pb collisions at 2.76 and 5.02 TeV, the parameters $A$, $B$ and $n$ have been obtained by logarithmic interpolations of the parameters $\ln A$, $B$ and $n$ between Au+Au collisions at 200 GeV and Pb+Pb collisions at 5.5 TeV \cite{Zhu:2014csa, Zhu:2021fbs}. Taking into account the centrality dependence, the minijets distribution is assumed as \cite{ Zhu:2021fbs}
\begin{eqnarray}
f_i(k, c) = \frac{T_{AA}(c)}{T_{AA}(0.05)}f_i(k, 0.05),
\label{211}
\end{eqnarray}
where $c=0.05$ stands for $0-10\%$ centrality. The values of nuclear thickness function $T_{AA}(c)$ for Pb+Pb collisions are available in Ref. \cite{Abelev:2013qoq}. Since $f_i(k, c)$ has a power-law dependence on $k$, so does $ \mathcal{S}^j(p, c)$ on $p$ in contrast to the exponential behavior of the thermal partons. The introduction of shower partons is the way to bring the effects of hard scattering to the hadronization scale in the recombination model. At the same time, the formalism does not exclude fragmentation by a hard parton, since SS recombination at high $p_T$ is equivalent to fragmentation. For charm quarks, we extract the initial distribution directly from PYTHIA8~\cite{Bierlich:2022pfr} generation using the Monash tune to derive the charm quark distribution in p+p collisions at $\sqrt{s_{NN}}=2.76$ and 5.02 TeV, respectively. Then the transverse momentum distribution of charm minijets at a given centrality class in Pb+Pb collisions from the corresponding distribution in p+p collisions can be computed, as suggested in Ref. \cite{dEnterria:2003xac}.

\section{transverse momentum distributions of charmed mesons}\label{DIS}
The formulae for recombination of thermal and shower partons were well developed for central collisions in heavy-ion collisions at RHIC and LHC \cite{hz2, Zhu:2014csa}. Up to now, we have successfully provided a coherent explanation for the centrality dependence of seven identified hadrons ($\pi$, p, K, $\Lambda$, $\phi$, $\Xi$, $\Omega$) produced in Pb+Pb collisions at $\sqrt{s_{NN}}=2.76$ and 5.02 TeV \cite{Zhu:2021fbs}. Now we turn our interest to the charmed mesons production, such as $D^0$, $D_s^+$ and $J/\psi$ using the quark recombination model.

\subsection{$D^0$ production}
Since the mass of charmed mesons are not negligible, we should replace $p_0$ in Eq. (\ref{21}) by the transverse mass $m_T=(m^2+p_T^2)^{1/2}$ at midrapidity. With the RF for $D^0$ \cite{Hwa:1994uha}
\begin{eqnarray}
R_{D^0}(p_1,p_2,p_T)={5p_1 p_2^5 \over p_T^5} \delta(p_1+p_2-p_T), \label{31}
\end{eqnarray}
the four components for $D^0$ production are,
\begin{eqnarray}
\frac{dN^{TT}_{D^0}}{p_Tdp_T}&=& \frac{5C_q C_c}{m_T^{D^0}p_T^6}\int_0^{p_T}dp_1 p_1e^{-p_1/T_q} \nonumber  \\ 
&&\times(p_T -p_1)^5 e^{-(p_T -p_1)/T_c}, 
\label{32}
\end{eqnarray}
\begin{eqnarray}
\frac{dN^{TS}_{D^0}}{p_Tdp_T}&=&{5\over m_T^{D^0} p_T^6}\int_0^{p_T}dp_1 (p_T -p_1)^4\nonumber \\
&&\times [p_1C_q e^{-p_1/T_q} S^c(p_T-p_1, c) \nonumber \\ 
&&\times+C_c(p_T -p_1) e^{-(p_T -p_1)/T_c}S^{\bar{u}}(p_1, c)] , \nonumber \\
\label{33}
\end{eqnarray}
\begin{eqnarray}
\frac{dN^{SS^{1j}}_{D^0}}{p_Tdp_T}&=&{1\over m_T^{D^0}}\int {dq \over q^2}\sum_{i=g,c} \hat{F_i}(q, c)D_i^{D^0}(p_T, q), 
\label{34}
\end{eqnarray}
\begin{eqnarray}
\frac{dN^{SS^{2j}}_{D^0}}{p_Tdp_T}&=&{5\Gamma \over m_T^{D^0}p_T^6}\int_{0}^{p_T}dp_1 (p_T-p_1)^4\nonumber \\ 
&&\times\mathcal{S}^{\bar{u}}(p_1, c) \mathcal{S}^{c}(p_T-p_1, c).
\label{35}
\end{eqnarray}

The shower-shower recombination from one jet ($SS^{1j}$) is equivalent to fragmentation, so the FF $D_i^{D^0}$ is directly used in Eq. (\ref{34}). $\Gamma$ in Eq. (\ref{35}) is the probability that two parallel partons originated from two jets can recombine. Same as done in Ref. \cite{Zhu:2014csa}, $\Gamma$ was estimated as 0.1.

\subsection{$D_s^+$ production}
The four components of $D_s^+$ are very similar to those of $D^0$. The differences are in the constituent quark masses between $m_q$ and $m_s$ and the different $T_q$ of light quarks and $T_s$ of $s$ quarks. The RF of $D_s^+$ is taken to be \cite{Hwa:1994uha}
\begin{eqnarray}
R_{D_s^+}(p_1,p_2,p_T)={3p_1 p_2^3 \over p_T^5} \delta(p_1+p_2-p_T). 
\label{36}
\end{eqnarray}
Therefore, we have the $D_s^+$ distributions,
\begin{eqnarray}
\frac{dN^{TT}_{D_s^+}}{p_Tdp_T}&=&{3C_s C_c\over m_T^{D_s^+} p_T^4}\int_0^{p_T}dp_1 p_1e^{-p_1/T_s}\nonumber \\ 
&&\times(p_T -p_1)^3 e^{-(p_T -p_1)/T_c}, 
\label{37}
\end{eqnarray}
\begin{eqnarray}
\frac{dN^{TS}_{D_s^+}}{p_Tdp_T}&=&{3 \over m_T^{D_s^+} p_T^{4}}\int_0^{p_T}dp_1 (p_T -p_1)^2\nonumber  \\ 
&&\times [C_s p_1e^{-p_1/T_s} S^c(p_T-p_1, c)\nonumber \\ 
&&+C_c(p_T-p_1)e^{-(p_T -p_1)/T_c}S^{\bar{s}}(p_1, c)], 
\label{38}
\end{eqnarray}
\begin{eqnarray}
\frac{dN^{SS^{1j}}_{D_s^+}}{p_Tdp_T}&=&{1\over m_T^{D_s^+}}\int {dq \over q^2}\sum_{i=g,c} \hat{F_i}(q, c)D_i^{D_s^+}(p_T, q),
\label{39}
\end{eqnarray} 
\begin{eqnarray}
\frac{dN^{SS^{2j}}_{D_s^+}}{p_Tdp_T}&=&{3\Gamma \over m_T^{D_s^+}p_T^{4}}\int_{0}^{p_T}dp_1(p_T-p_1)^2\nonumber \\
&&\times\mathcal{S}^{\bar{s}}(p_1, c) \mathcal{S}^{c}(p_T-p_1, c).
\label{310}
\end{eqnarray}

\subsection{$J/\psi$ production}
For $J/\psi$,  $m_c =m_{\bar c} =1.5$ GeV and $m_{J/\psi}\approx3.1$ GeV. The relationship between the masses imply that the interaction between $c$ and $\bar c$ is very weak. Considering the fact that $c$ and $\bar c$ are moving together with the same velocity as $J/\psi$, the RF of $J/\psi$ can be written as
\begin{eqnarray}
R_{J/\psi}(p_1,p_2,p_T)=g_{J/\psi}p_1p_2\prod_{i=1}^2 \delta(p_i-p_T/2), 
\label{311}
\end{eqnarray}
with the statistical factor $g_{J/\psi}$. Therefore, the four components for $J/\psi$ are
\begin{eqnarray}
\frac{dN^{TT}_{J/\psi}}{p_Tdp_T}&=&{g_{J/\psi}C_c^2 p_T \over 4m_T^{J/\psi}}e^{-p_T/T_c}, 
\label{312}
\end{eqnarray}
\begin{eqnarray}
\frac{dN^{TS}_{J/\psi}}{p_Tdp_T}&=&{g_{J/\psi}C_c \over 2m_T^{J/\psi}}e^{-p_T/2T_c}\mathcal{S}^c(p_T/2, c),
\label{313}
\end{eqnarray}
\begin{eqnarray}	
\frac{dN^{SS^{1j}}_{J/\psi}}{p_Tdp_T}&=&{1 \over p_T m_T^{J/\psi}}\int {dq \over q^2}\sum_{i=g,c} \hat{F_i}(q, c)D_i^{J/\psi}(p_T, q),\nonumber\\
\label{314}
\end{eqnarray}
\begin{eqnarray}
\frac{dN^{SS^{2j}}_{J/\psi}}{p_Tdp_T}&=&{g_{J/\psi}\Gamma \over p_T m_T^{J/\psi}}\mathcal{S}^c(p_T/2, c)\mathcal{S}^{\bar{c}}(p_T/2, c).
\label{315}
\end{eqnarray}

 \begin{figure}[pht]
\includegraphics[width=0.45\textwidth]{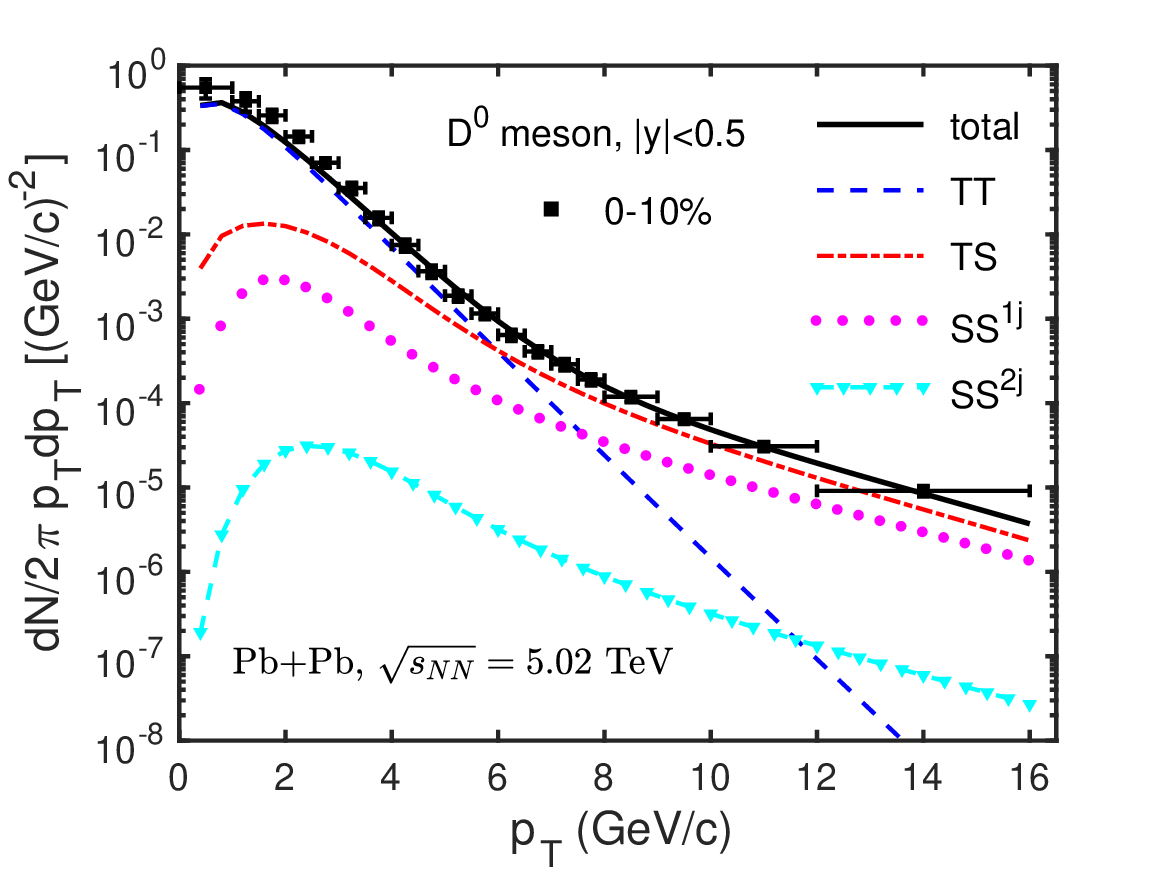}
 \caption{(Color online) Transverse momentum spectrum of $D^0$ at the centrality of 0-10\% in Pb+Pb collisions at $\sqrt{s_{NN}}=5.02$ TeV. The data are taken from Ref. \cite{ALICE:2021rxa}.}
 \label{fig1}
\end{figure}

\begin{figure}[pht]
\includegraphics[width=0.45\textwidth]{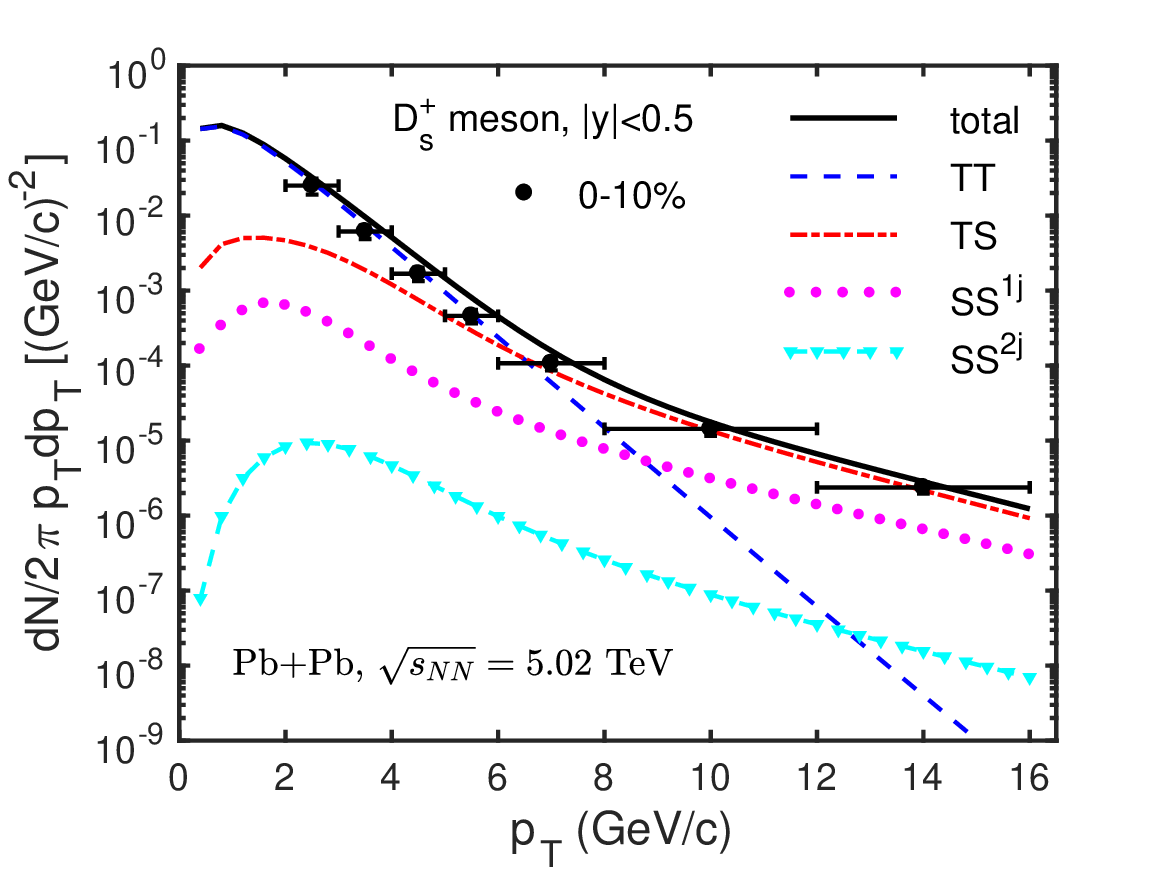}
 \caption{(Color online) Transverse momentum spectrum of $D_s^+$ at the centrality of 0-10\% in Pb+Pb collisions at $\sqrt{s_{NN}}=5.02$ TeV. The data are taken from Ref. \cite{ALICE:2021kfc}.}
 \label{fig2}
\end{figure}

\begin{table*}
\tabcolsep0.2in
\begin{tabular}{ccccc}
\hline
 $\sqrt{s_{NN}}$ [TeV] & centrality  &  $C_q$ [(GeV/c)$^{-1}$]& $C_s$  [(GeV/c)$^{-1}$]& $C_c$  [(GeV/c)$^{-1}$]\\ 
 \hline
  &0-10\%  &22.4 &10.6 &2.9\\
2.76&20-40\% &15.5 &7.3 &0.9\\
&20-50\%  &14.3 &6.7 &0.9\\
&30-50\%  &13.2 &6.2 &0.9\\
 \hline
 5.02&0-10\%  &23.8 &8.0&1.1 \\
&30-50\% &15.0 &4.0 &0.39\\
 \hline
 \end{tabular}
 \caption{Parameters $C_q$, $C_s$ and $C_c$ for Pb+Pb collisions at $\sqrt{s_{NN}}=2.76$ and 5.02 TeV, respectively.} 
 \label{tab3}
 \end{table*}
 
\section{results and discussion}\label{results}
In the present section, we show the results on the transverse momentum spectra and the second coefficient of azimuthal anisotropy $v_2$ for $D^0$, $D_s^+$ and $J/\psi$ in Pb+Pb collisions at $\sqrt{s_{NN}}=2.76$ TeV and 5.02 TeV. The inverse slopes $T_q$ and $T_s$, the values of $\gamma_0$ and $q_0$ in Eq. (\ref{29}) are shown in Tables \ref{tab1} and \ref{tab2}, respectively. Furthermore, the normalization factor $C_q$ of $u, d$ quarks and $C_s$ of strange quarks for Pb+Pb collisions at $\sqrt{s_{NN}}=2.76$ TeV and 5.02 TeV have been obtained in Ref. \cite{Zhu:2021fbs} and are listed in TABLE \ref{tab3}. Therefore, the unknown parameters are $T_c$ and the centrality dependence of normalization factor $C_c$ for charm quarks.

\begin{figure}[pht]
\includegraphics[width=0.45\textwidth]{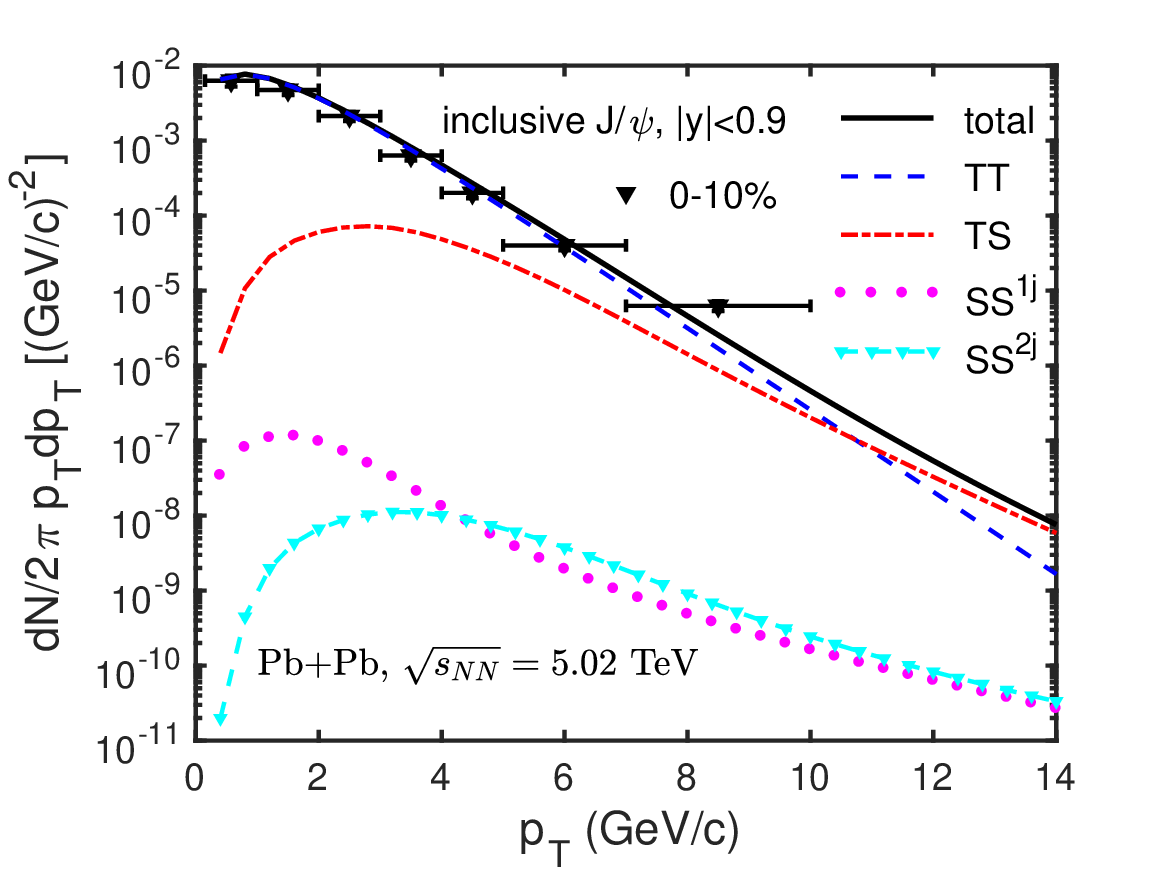}
 \caption{(Color online) Transverse momentum spectrum of $J/\psi$ at the centrality of 0-10\% in Pb+Pb collisions at $\sqrt{s_{NN}}=5.02$ TeV. The data are taken from Ref. \cite{ALICE:2023gco}.}
 \label{fig3}
\end{figure}

 \begin{figure*}[t]
 \centering
\includegraphics[width=0.7\textwidth]{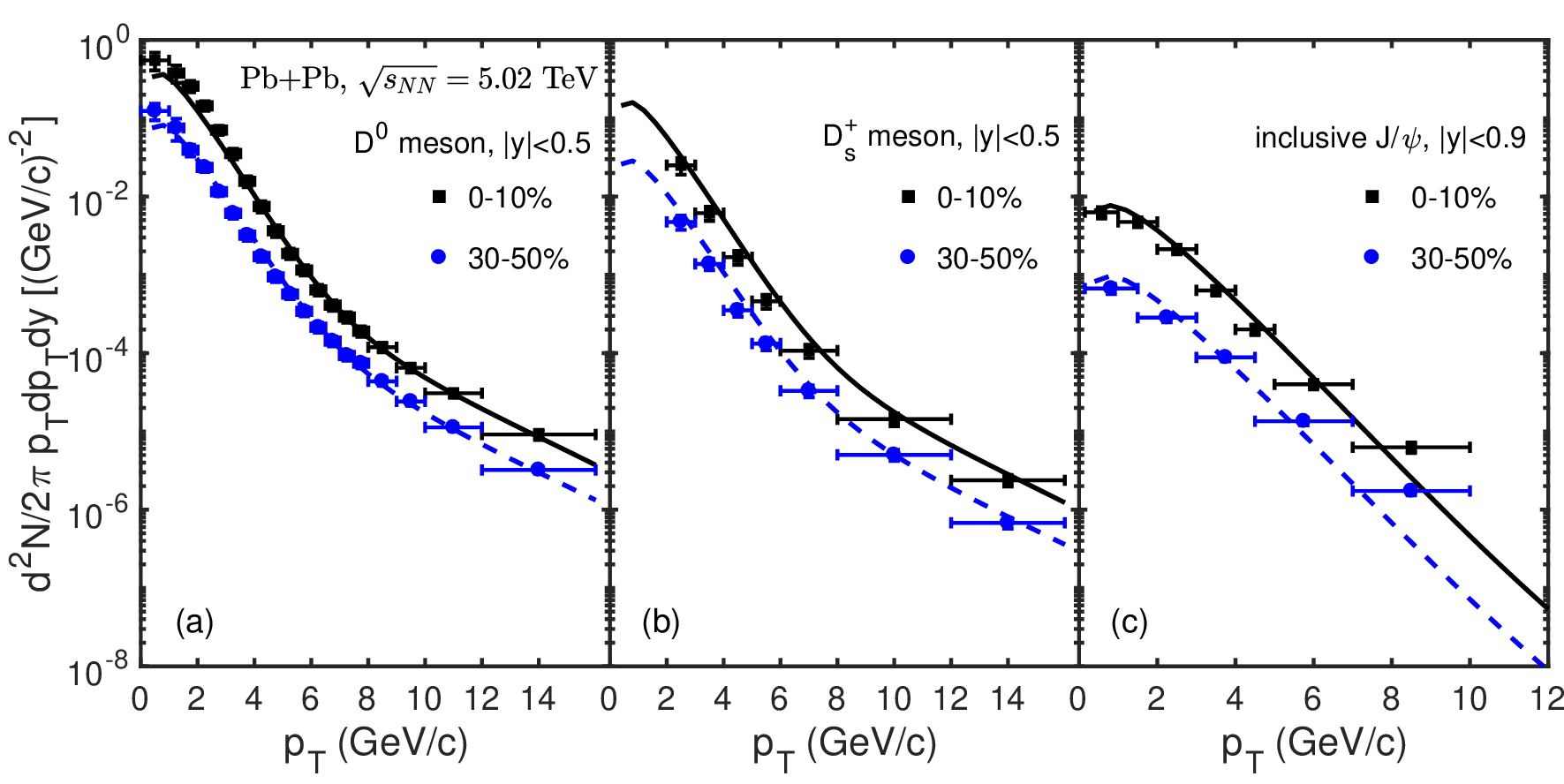}
  \caption{(Color online) Transverse momentum spectra of $D^0$ (a), $D_s^+$ (b) and $J/\psi$ (c) from the recombination model at midrapidity in Pb+Pb collisions at $\sqrt{s_{NN}}=5.02$ TeV for different centrality classes. The data are taken from Ref. \cite{ALICE:2021rxa} for $D^0$, Ref. \cite{ALICE:2021kfc} for $D_s^+$ and Ref. \cite{ALICE:2023gco} for $J/\psi$.}
\label{fig4}
\end{figure*}

\subsection{Transverse momentum spectra}
In Fig. \ref{fig1}, we show the transverse momentum spectrum of $D^0$ (the black solid line) produced in Pb+Pb collisions at $\sqrt{s_{NN}}=5.02$ TeV and centrality of 0-10\% at midrapidity. Also shown in the figure are experimental $D^0$ spectrum from the ALICE Collaboration at centrality 0-10\% by black solid squares \cite{ALICE:2015vxz}. It is seen that the recombination model can nicely describes the experimental data. The thermal and shower partons in various combinations are shown by different line types, which dominate at different $p_T$ regions. The jet-medium interaction is taken into account at the hadronization stage by TS component, which dominates at intermediate $p_T$.  SS recombination from one jet becomes important at high $p_T$, which is equivalent to fragmentation but in a language that has dynamical content at the hadronization scale. The value of  parameter $\gamma_0$ in $\gamma$ factor for charm quarks is 1.0 for $\sqrt{s_{NN}}=5.02$ TeV, which is smaller than that for light quarks. It means that the effect of jet quenching in the medium for charm quarks is smaller than that for light quarks. In contrast to light quarks and gluons, which can be produced or annihilated during the entire evolution of the medium, heavy quarks are produced in the initial hard-scattering processes and their annihilation rate is small \cite{Braun-Munzinger:2007fth}. Therefore, heavy quarks preserve their flavor and mass identity while traversing the medium and can be tagged throughout all momentum ranges, from low to high $p_T$, through the measurement of heavy flavor hadrons in the final state of the collision.  The inverse slope for charm quarks $T_c$ is determined by fitting the transverse momentum spectrum of $D^0$ at low $p_T$, which is listed in TABLE \ref{tab1}. The values of $C_c$ are given in TABLE \ref{tab3}. Similar result is obtained for the transverse momentum spectrum of $D_s^+$ in central Pb+Pb collisions at $\sqrt{s_{NN}}=5.02$ TeV. The agreement between the calculation results from RM and experimental data is quite good, as shown in Fig. \ref{fig2}. 

 \begin{figure*}[t]
 \centering
\includegraphics[width=0.7\textwidth]{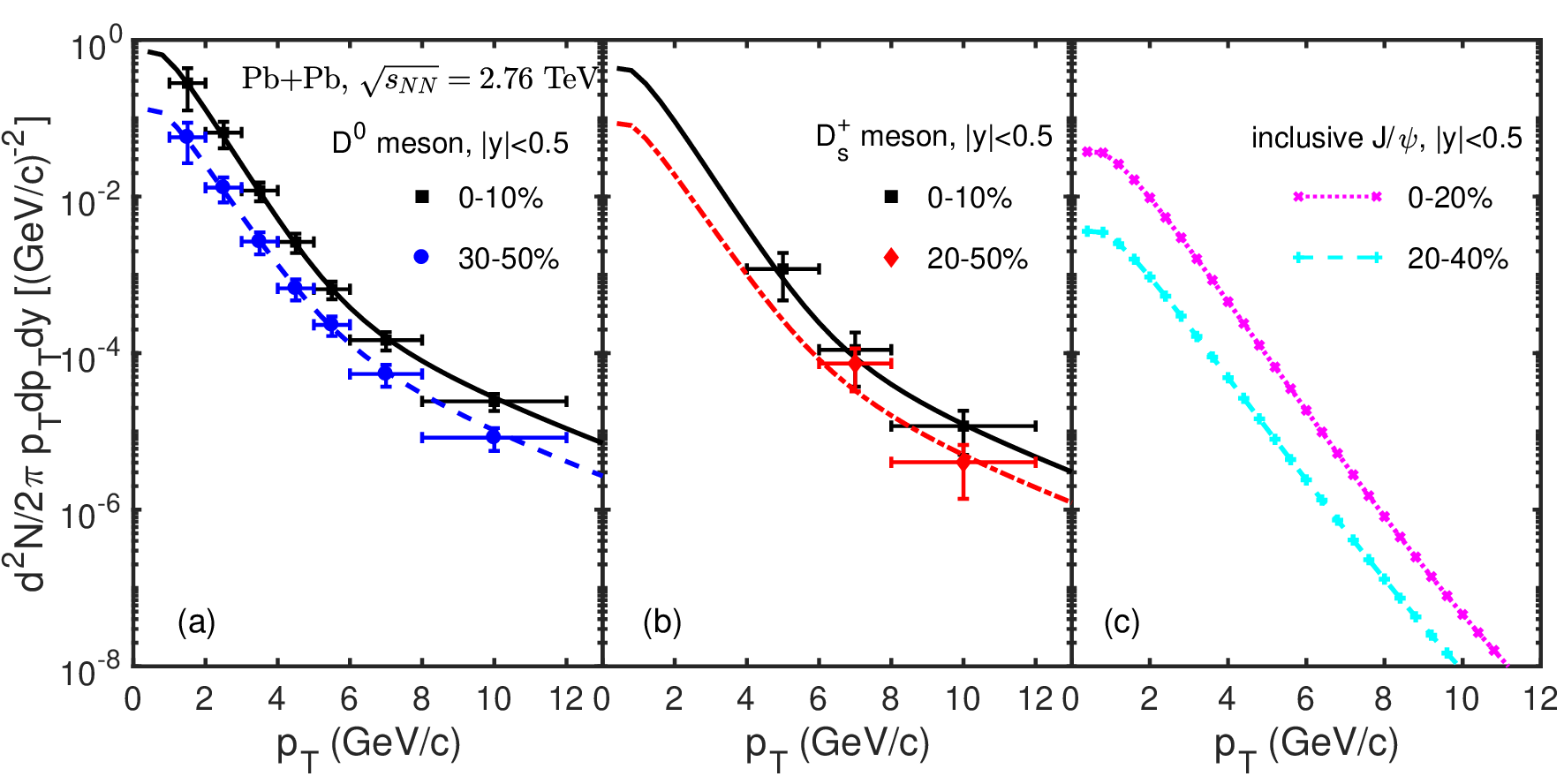}
  \caption{(Color online) Transverse momentum spectra of $D^0$ (a), $D_s^+$ (b) and $J/\psi$ (c) from the recombination model at midrapidity in Pb+Pb collisions at $\sqrt{s_{NN}}=2.76$ TeV for different centrality classes. The data are taken from Ref. \cite{ALICE:2015vxz} for $D^0$ and Ref. \cite{ALICE:2015dry} for $D_s^+$. }
\label{fig5}
\end{figure*}

Figure \ref{fig3} shows the calculation result for $J/\psi$ at the centrality of 0-10\% in Pb+Pb collisions at $\sqrt{s_{NN}}=5.02$ TeV. What is most noticeable is that the relative strengths of the components are unlike the cases for $D^0$ and $D_s^+$. The TT component dominates for $p_T<7$ GeV/c and that SS component is much lower. It means that $J/\psi$ is produced thermally even at $p_T\sim7$ GeV/c without any contribution from parton fragmentation which is the usual mechanism considered in pQCD. This result is understandable. At earlier time $J/\psi$ is formed at higher density, its survival in the hot medium is suppressed due to its dissociation through interaction with the plasma that is still active.  By varying $g_{J/\psi}$ only for the overall normalization, we obtain the good description for $J/\psi$ spectrum at the centrality of 0-10\% with $g_{J/\psi}=0.28$. The small value of $g_{J/\psi}$ relative to $g_{D^0}$ and $g_{D_s^+}$ being 1 is an indication of charmonium suppression after $J/\psi$ is formed at a time much earlier than $D^0$ and $D_s^+$, when the density of charm quarks is higher. Like the case of $\phi$ meson production discussed in our earlier work \cite{Zhu:2014csa},  $J/\psi$ also experiences the effects of dissociation by the plasma as it traverses the remaining portion of the medium before it completely hadronizes.

Figure \ref{fig4} includes the calculation results for the transverse momentum spectra of  $D^0$, $D_s^+$ and $J/\psi$ for the centralities of 0-10\% and 30-50\% at midrapidity in Pb+Pb collisions at $\sqrt{s_{NN}}=5.02$ TeV. Evidently, for $D^0$, $D_s^+$  and $J/\psi$ the agreement with data for both central and non-central collisions is excellent for all $p_T$ where data exist. After investigating the charmed mesons production in Pb+Pb collisions at $\sqrt{s_{NN}}=5.02$ TeV, it is natural to apply it to the lower colliding energy  $\sqrt{s_{NN}}=2.76$ TeV. The inverse slope $T_c$ is determined by the $D^0$ distribution at central collisions shown in TABLE \ref{tab1}. With the retuned parameters $C_c$,  our calculation results can nicely reproduce the transverse momentum spectra of $D^0$ and $D_s^+$  for various centralities at $\sqrt{s_{NN}}=2.76$ TeV. The value of parameter $\gamma_0$ in $\gamma$ factor is 1.2, which is a little bit larger than that for  $\sqrt{s_{NN}}=5.02$ TeV. It is remarkable that the calculated distributions agree well with the experimental data of charmed mesons for both colliding energies at LHC. 
Unfortunately, the experimental data for $J/\psi$ at midrapidity in Pb+Pb collisions at $\sqrt{s_{NN}}=2.76$ TeV are not available yet. We predict the transverse momentum distributions of $J/\psi$ at the centralities of 0-20\% and 20-40\% in Fig. \ref{fig5}c, which can be compared with experimental measurements in the near future.  
From the above study, we can conclude that our results from RM agree well with the measured data at all centralities and transverse momenta for the light hadrons and charmed mesons, which gives support to the reliability of the dynamical roles that minijets and their shower partons play in our model.

\subsection{Second harmonic of azimuthal anisotropy} 
We now broaden our consideration to include the azimuthal angle $\phi$ dependence by considering the effects of semihard partons, which are created near the surface and directed outward. They can generate the minijets and give rise to $\phi$ anisotropy in the thermal component.  As done in Ref. \cite{Hwa:2012xy}, $\rho^h(p_T, \phi, b)$ denotes the single-particle distribution of hadron $h$ produced at mid-rapidity in heavy-ion collisions at impact parameter $b$
\begin{eqnarray}
\rho^h(p_T,\phi, b) = {dN_h\over p_Tdp_Td\phi}(p_T, \phi, b).     \label{41}
\end{eqnarray}
At low $\pt$, $\rho^h$ is assumed to consist of three components
\bq
\rho^h(p_T,\phi, b) =B^h(p_T, b) + R^h(p_T,\phi, b) + M^h(p_T,\phi, b),   \nonumber \\   \label{42}
\eq
referring as Base, Ridge and Minijet components, respectively.
The first two components are due to the recombination of thermal partons (TT), while the third one is dependent of thermal-shower (TS) for mesons, which is dominant in the intermediate $p_T$ region, but is still not negligible at low $p_T$. 
Therefore, the recombination of thermal partons has two components.
One is the base component, which is azimuthally isotropic, while the other one is the ridge, which is $\phi$ dependent. On the other hand, in heavy-ion collisions, the density of minijets produced by semihard scattering of partons is so high that it is important to consider how they affect the azimuthal harmonics.
After averaging over $\phi$, we have
\bq
\bar\rho^h(p_T, b) =B^h(p_T,  b) +\bar{R}^h(p_T, b) + \bar{M}^h(p_T, b).     \label{43}
\eq
With Eq.\ (\ref{42}), the second harmonic coefficient can be expressed as
\begin{eqnarray}
v_2^h(p_T,b) = \langle \cos n\phi \rangle_{\rho}^h = {\int_0^{2\pi} d\phi \cos 2\phi\rho^h(p_T,\phi,b)\over \int_0^{2\pi} d\phi\rho^h(p_T,\phi,b)}.     \label{44}
\end{eqnarray}

 \begin{figure}[pht]
\includegraphics[width=0.45\textwidth]{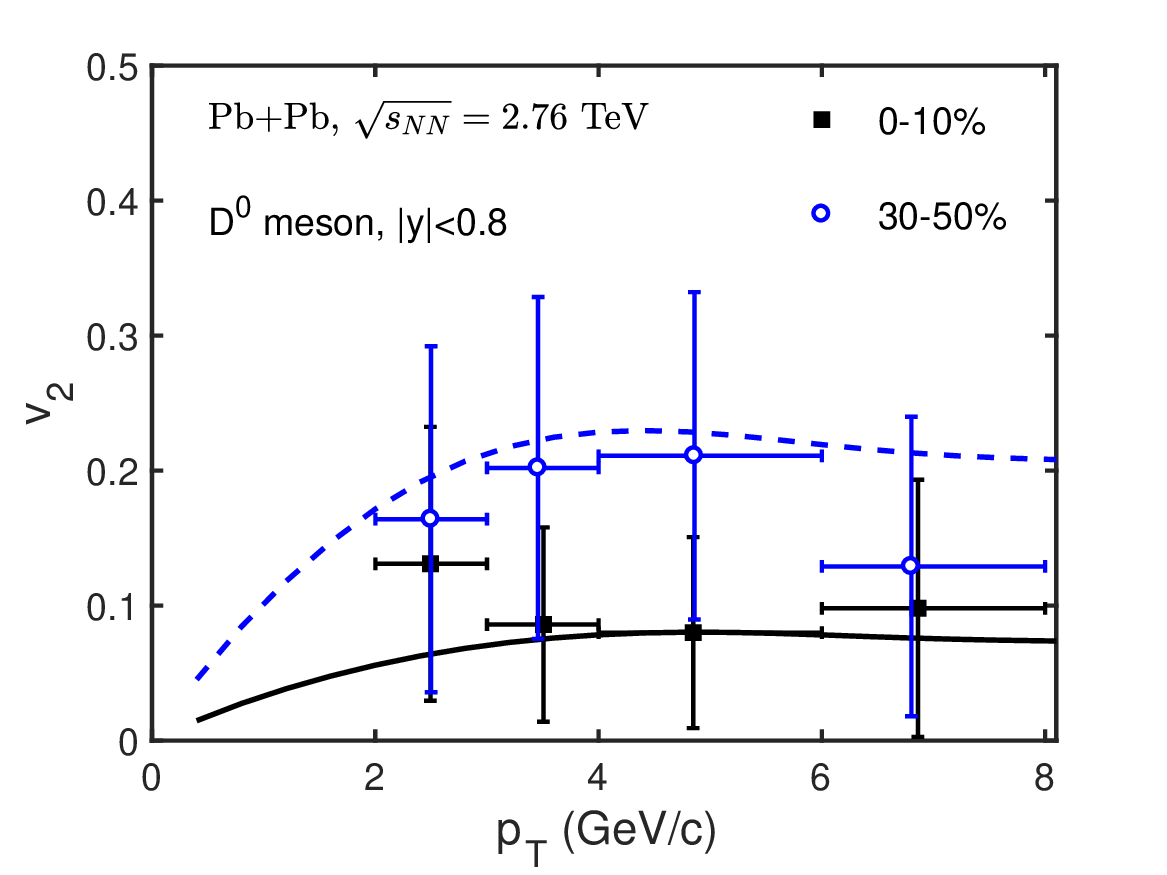}
 \caption{(Color online) The second harmonic coefficient $v_2$ of $D^0$ at the centralities of 0-10\% and 30-50\% in Pb+Pb collisions at $\sqrt{s_{NN}}=2.76$ TeV. The data are taken from Ref. \cite{ALICE:2014qvj}.}
 \label{fig6}
\end{figure}

 \begin{figure*}[t]
 \centering
\includegraphics[width=0.75\textwidth]{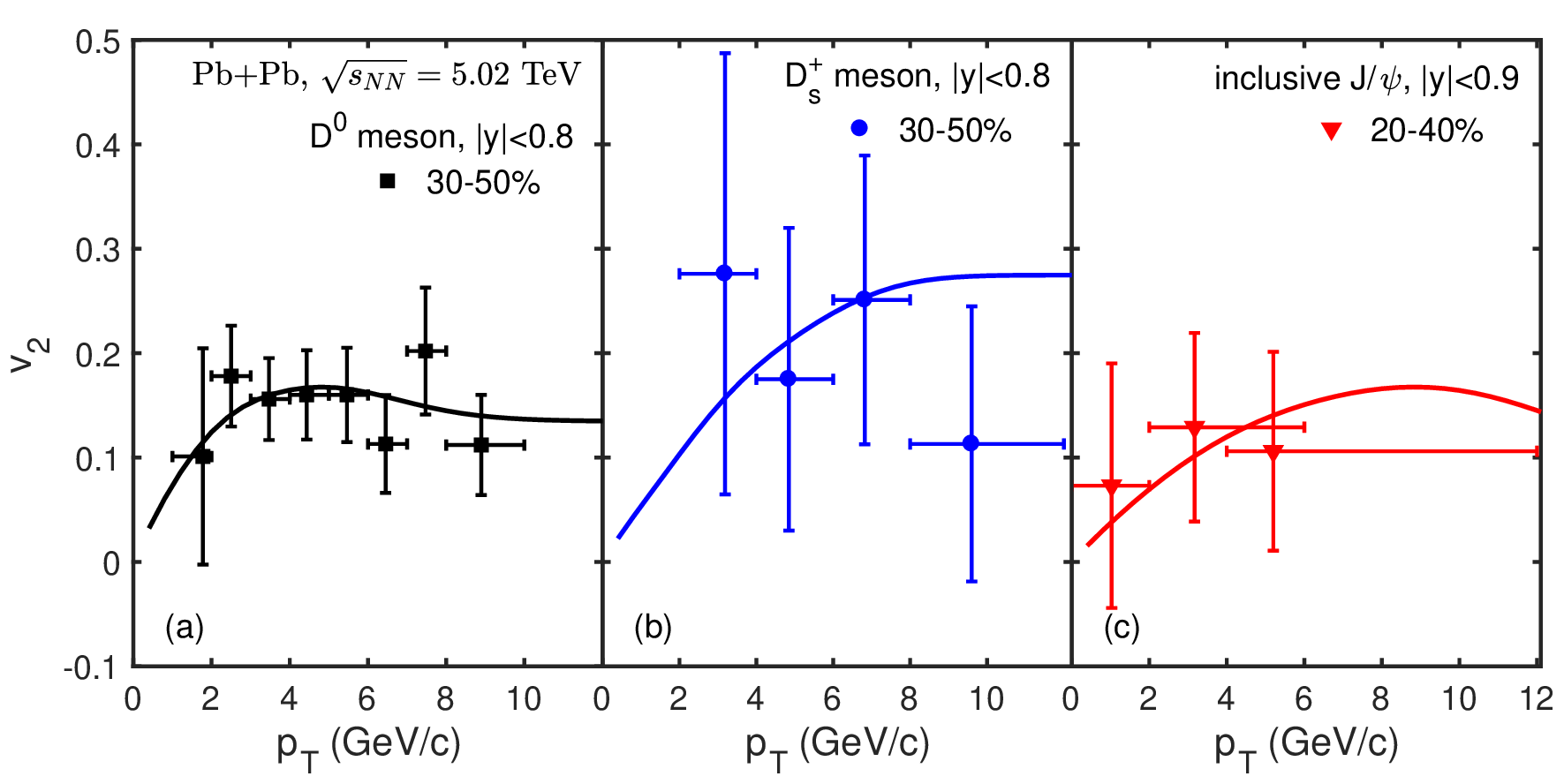}
  \caption{(Color online) The second harmonic coefficient $v_2$ of $D^0$ (a), $D_s^+$ (b) and $J/\psi$ (c) as a function of transverse momentum in Pb+Pb collisions at $\sqrt{s_{NN}}=5.02$ TeV for different centrality classes. The data are taken from Ref. \cite{ALICE:2017pbx} for $D^0$, Ref. \cite{ALICE:2021kfc} for $D_s^+$ and Ref. \cite{ALICE:2017quq} for $J/\psi$.}
\label{fig7}
\end{figure*}

The $\phi$-independent base thermal component is expressed as
\bq
B^h(p_T, b) ={\cal N}_h(p_T, b)e^{-\pt/T_0},  \label{45}
\eq
where ${\cal N}_h(p_T, b)$ is the normalization factor. $T_0$ is treated as a free parameter, which is colliding energy dependent.
The second component $R(p_T, \phi, b)$ contains the $\phi$ dependence due to the initial elliptical spatial configuration through the $S(\pb)$. The physical meaning of $S(\pb)$ is that it is the segment on the initial ellipse through which semihard partons are emitted to contribute to a ridge particle at $\phi$.  In Ref. \cite{Chiu:2008ht}, it was found that the azimuthal correlation between a semihard parton and a ridge particle formed by TT recombination can be described by a Gaussian distribution in $\phi$ with a width $\sigma=0.33$ in order to reproduce the ridge data. Therefore, the possible semihard partons that can contribute to the fixed ridge particles at $\phi$ must have their azimuthal angles that deviate from $\phi$ by no more than $\sigma$ \cite{Chiu:2008ht, Hwa:2012xy}. This means the exit point of semihard parton at the surface is restricted to a certain range $\phi\pm \sigma$. 
The derivation of $S(\phi,b)$ given in Ref. \cite{Hwa:2009rd} is based on the geometry of the initial configuration taken to be an ellipse with width $w$ and height $h$, where $w=1-b/2$ and $h=(1-b^2/4)^{1/2}$ in units of nuclear radius $R_A$,
 \begin{eqnarray}
S(\phi,b) = h(b)[E(\theta_2,\alpha) - E(\theta_1,\alpha)],     \label{46}
\end{eqnarray}
where $E(\theta_i,\alpha)$ is the elliptic integral of the second kind with $\alpha=1-w^2/h^2$ and
\begin{eqnarray}
\theta_i = \tan^{-1} \left({h\over w}\tan\phi_i \right), \quad \phi_1 = \phi - \sigma, \quad \phi_2 = \phi + \sigma.  \nonumber \\    \label{47}
\end{eqnarray}
It should be noted that $S(\phi,b)$ in Eq. (\ref{46}) is not normalized. Since the elliptical axes may not coincide with the reaction plane that contains the impact parameter $b$, a tilt angle $\psi_2$ is introduced and averaged over,
\begin{eqnarray}
\tilde S_2(\phi, b)=\frac{2}{\pi}\int_{\pi/4}^{\pi/4}d\psi_2 S_2(\phi-\psi_2, b).     \label{48}
\end{eqnarray}
So the normalized $\tilde S_2(\phi,b)$ can be expressed as,
\begin{eqnarray}
S_2(\phi, b)=\tilde S_2(\phi, b) \left /{\frac{1}{2\pi}\int_{0}^{2\pi}d\phi \tilde S_2(\phi, b)}\right. .    \label{49}
\end{eqnarray}
Then, the ridge component in Eq. (\ref{42}) which responds to the minijets through TT recombination can be written as
\begin{eqnarray}
R^h(\ppb)=S(\pb) \bar R^h(p_T, b),  \label{410}
\end{eqnarray}
where $ \bar R^h(p_T, b)$ is the second component of recombination of thermal partons ($\frac{dN^{TT}_h}{p_Tdp_T}$). The first component is the $\phi$-independent $B^h(p_T, b)$. As done in Ref. \cite{Hwa:2012xy}, we write the third component of $\rho^h(\ppb)$ as
\begin{eqnarray}
M^h(\ppb)=J(\pb) \bar M^h(\pt,b), \label{411}
\end{eqnarray}
where $\bar M^h(\pt,b)$ refers to the recombination of thermal-shower partons ($\frac{dN^{TS}_h}{p_Tdp_T}$), which is given in Sec. \ref{DIS}.  $J(\pb)$ describes the $\phi$-dependent part of the minijet contribution, which is assumed as 
\begin{eqnarray}
J(\pb)=\tilde J(\pb) \left /{\frac{1}{2\pi}\int_{0}^{2\pi}d\phi \tilde J(\pb)}\right. ,  \label{412}
\end{eqnarray}
where $\tilde J(\pb)$ contains all the harmonic components, $\cos n\phi$, averaged over the tilt anlge $\psi_n$,
\begin{eqnarray}
\tilde J(\pb)=1+b \sum_{n=2}^\infty a_n {n\over \pi}\int_{-\pi/2n}^{\pi/2n} d\psi_n \cos n(\phi-\psi_n),  \label{413}
\end{eqnarray}
where $a_n, n\ge 2$, are free parameters to be determined by fitting the experimental data of $v_n$.

 \begin{table}
\tabcolsep0.3in
\renewcommand{\arraystretch}{1.3}
\begin{tabular}{ccc}
\hline
\hline
 $\sqrt{s_{NN}}$ (TeV) &2.76 &5.02\\ 
 \hline
$D^0$  & 0.46 & 0.3 \\
$D_s^+$  &&0.61 \\
$J/\psi$ &&0.21 \\
 \hline
  \hline
 \end{tabular}
 \caption{Parameter $a_2$ in Eq. (\ref{413}) for $D^0$, $D_s^+$ and $J/\psi$ in Pb+Pb collisions at $\sqrt{s_{NN}}=2.76$ and 5.02 TeV, respectively.} 
 \label{tab4}
 \end{table}

With the three components of $\rho^h(\ppb)$ shown in  Eqs.\ (\ref{45}), (\ref{410}) and (\ref{411}),  a universal formula for the second azimuthal harmonic coefficient from Eq.\ (\ref{41}) can be obtained
\begin{eqnarray}
v_2^h(\pt,b)={\left<\cos 2\phi\right>_S \bar R^h(\pt,b) + \left< \cos 2\phi\right>_J \bar M^h(\pt,b) \over \bar\rho^h(\pt,b)}  , \nonumber \\  \label{414}
\end{eqnarray}
where
\begin{eqnarray}
\langle \cos2\phi \rangle_S &=& {1\over 2\pi} \int_0^{2\pi} d\phi \cos2\phi S(\phi,b),     \label{414} \\
\left< \cos 2\phi\right>_J&=&{1\over 2\pi} \int_0^{2\pi} d\phi \cos 2\phi J(\pb).  \label{415}
\end{eqnarray}

In Fig. \ref{fig6}, we show the second harmonic coefficient $v_2$ of $D^0$ at the centralities of 0-10\% and 30-50\% in Pb+Pb collisions at $\sqrt{s_{NN}}=2.76$ TeV from the recombination model. Compared to the experimental data, shown by black full squares for 0-10\% and blue empty circles for 30-50\%, from the ALICE Collaboration \cite{ALICE:2014qvj}.  The similar results are also obtained for $D^0$, $D_s^+$ and $J/\psi$ produced at midrapidity in Pb+Pb collisions at $\sqrt{s_{NN}}=5.02$ TeV, as shown in Fig. \ref{fig7}. The solid lines reproduce very well the data from ALICE Collaboration \cite{ALICE:2017pbx, ALICE:2021kfc, ALICE:2017quq} at various centralities. It is remarkable that the calculated curves agree well with the data for $p_T$ almost up to 10 GeV/c. The parameter $a_2$ is independent of the centrality, whose values are shown in TABLE \ref{tab4}.

\section{summary}\label{summary}
In this paper,  we have studied the transverse momentum distributions and azimuthal anisotropy of charmed mesons in relativistic heavy-ion collisions at LHC within the framework of recombination model that includes the effects of minijets, which can generate azimuthal anisotropy both through energy loss to the medium and in creating shower partons that recombine with the thermal partons. With emphasis on the effects of minijets, calculated results agree well with the experimental data for $p_T$ spectra and the second harmonic $v_2$ of $D^0$, $D_s^+$ and $J/\psi$ in Pb+Pb collisions at $\sqrt{s_{NN}}=2.76$ and 5.02 TeV. 
The geometry and nuclear medium produced in heavy-ion collisions are complex, but the agreement between our model calculations and available data indicates that the current recombination model supports the picture that the centrality dependence of light hadrons and charmed mesons production for the whole $p_T$ region can be described by the recombination of thermal and shower partons in relativistic heavy-ion collisions. In this case, it is of great interest to study the production of $\Lambda_c^+$ baryons with the recombination model which we will discuss in a following paper.

\section{Acknowledgements}
This work was supported by the Natural Science Foundation of Sichuan Province under Grant No. 2023NSFSC1322 and the National Natural Science Foundation of China under Grant No. 11905120. H.Z. acknowledges the financial support from Key Laboratory of Quark and Lepton Physics in Central China Normal University under grant No. QLPL2024P01.


\end{document}